# Proposal on the Calculation of the Ionisation-Cluster Size Distribution

## (I). The Model and Its Simulation Methodology

**(B. Heide, Karlsruhe Institute of Technology (KIT), Institute for Nuclear Waste Disposal (INE), Bernd.Heide@kit.edu)**

## Abstract

**A statistical model (within the framework of the canonical approximation) is proposed for the calculation of the ionisation-cluster size distribution in nanodosimetry. It is inspired by the maximum entropy principle and the well-known nuclear droplet model. The model allows statements about thermodynamic state quantities and it can be applied in particular to the scenario of "low-energy (< 100 eV) electrons moving in nanovolumes (in the order of a few nanometers)" – a scenario for which trajectory models should not be used. The focus of this work is on presenting the model.**

## 1. Introduction

In the macroscopic range, i. e. for volumes with a size of millimetres or more, the absorbed dose (**$D_{abs}$**) and the linear energy transfer are good quantities to describe the action of ionising radiation in soft matter. Although $D_{abs}$ is a point quantity, however, it is often not sufficient for the microscopic (µm) length scale. This is because $D_{abs}$ refers to the average absorbed energy and hence deviations from the average absorbed energy are not visible in the absorbed dose. But deviations from the average value, which can become relatively large for volumes in the range of micrometres, are important since a single ionisation (i. e. energy deposition) can decompose a critical molecule and cause serious radiation damage (1). In microdosimetry, $D_{abs}$ is therefore replaced by the specific energy, which is a stochastic quantity contrary to $D_{abs}$ which is a statistical one. Likewise, linear energy transfer in microdosimetry, which refers to a mean (!) energy loss and can be understood as a "force" – in the general sense, is replaced by a stochastic quantity: lineal energy.

When considering even smaller volumes, i. e. nanometric volumes, it is found that the number of ion pairs generated in a gas detector (which simulates a nanometric target by applying a very low gas density) is no longer proportional to the deposited energy. In nanodosimetry, the quantities "specific energy" and "lineal energy" are therefore no longer in the spotlight. Instead, the focus is on the ionisations generated in the target only. The collectivity of ionisations generated in the target is denoted as "ionisation cluster (**ic**)"; their count is referred to as "size (**s**)". The action of ionising radiation on biological matter within nanometric volumes is now estimated based on the ionisation-cluster size distribution.

There are several models describing biological damage once a cell was exposed to ionising radiation (2 – 11). Except for the model of W. Friedland et al. (3), S. A. Ngcezu and H. Rabus (7), B. Grosswendt (8, 9) as well as the model of F. Villegas et al. (10), the models already start from an intial ionisation (energy deposition) aggregation or even from deoxyribonucleic acid (**DNA**) single-strand breaks (**SSBs**) and double-strand breaks (**DSBs**). However, it is in general not explicitly stated how the initial ionisation aggregation is transformed into clusters of ionisations and DNA breaks (the interaction points need not be equal with the break locations due to rearrangement processes; cf. also section "2. Model"), respectively. On the other hand, the models that deal with clustering contain at least one "free" parameter, i. e. a parameter for which there may be a relatively large margin of error (such as the cluster distance (10); it causes preselected ionisation-cluster sizes to some extent). In addition, all of the above models have one major disadvantage in common: their theoretical descriptions are based on classical trajectories. And this generally makes little sense for low-energy electrons or positrons (i. e. light, charged, low-energy leptons in general) that move in nanovolumes as the spatial extension of their wave packet is usually larger than the target volume itself! For example, if an initial electron is expected to move in a cube of 2 nm in dimension and has an energy of 100 (± 0.4) eV, i. e. a "well-defined" momentum with a relative uncertainty of 0.2%, then Heisenberg's uncertainty principle gives a spatial extension of the electron wave packet of about 60 nm (= 30 times the cube length). An approximate illustration of this situation is depicted in Figure 1. The "0.4 eV" uncertainty of the energy is achievable with a modern control unit for the "Kimball Physics ELG-2A Electron Gun", for instance (27). In addition to the large lack of knowledge about the ionisation positions, which are absolutely necessary for cluster formation in the previous models, a comparison of the de Broglie wavelength with the mean distance of the molecules in the target volume also shows that a classical trajectory approach is not adequate. Assuming that the target contains liquid water (water generally is the medium of choice for dosimetric calculations) and has a temperature of 37 °C, the average distance between the molecules is 0.311 nm (12). However, the de Broglie wavelength of 100 eV electrons is 0.123 nm, which is of the order of 0.311 nm, so quantum mechanical effects are relevant. (Side note: though this does not apply to 100 eV protons, which have a much smaller de Broglie wavelength of 0.003 nm due to their mass.) The methods used so far, which are based on classical trajectories, are therefore fundamentally not suitable for low-energy electrons. Quantum mechanical methods, on the other hand, are very sophisticated and enormously computationally intensive. To further address the clustering of ionizations caused by low-energy electrons, we propose a statistical model (which could, in principle, be extended using quantum statistics). This model is based, among other things, on the circumstance that the sheer number of ionisation interactions should not be affected by Heisenberg's uncertainty principle (cf. Section 2 "Circumstantial validity of the trajectory method" in (28)).

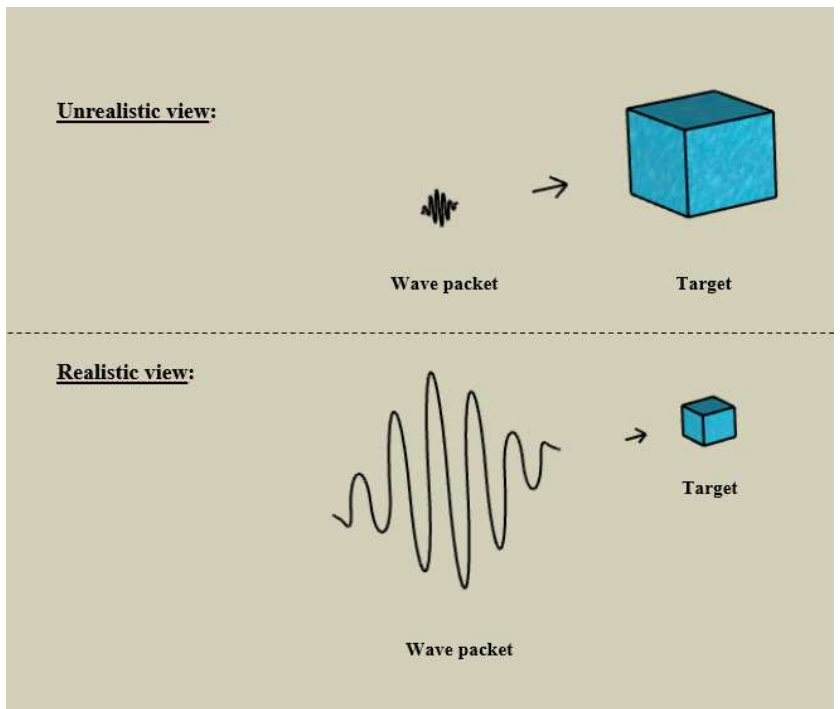

*Figure 1: approximate illustration of the proportions between the wave packet and the cube-shaped target. Upper half: unrealistic view (classical). Lower half: realistic view (quantum mechanical).*

In the following, we
(a) propose a ('semi-classical' in a sense) statistical model for the calculation of the ionisation-cluster size (**ics**) distribution in nanodosimetry which is detached from exact location information and uses as input only the total number of ionisations within a target volume, that can be determined by a Monte Carlo particle transport code (such as the program package Geant4-DNA (13, 14, 24 – 26)) and, hence, has no free parameter, and
(b) depict its simulation methodology in detail.
The model is based on a canonical ensemble. The method is inspired by the maximum entropy principle (29) and the well-known nuclear droplet model (19, 20), which has already been applied successfully for the prediction of nuclear-cluster distributions. The model can be applied in particular to a so-called "transition zone" for which the classical description using trajectories is no longer adequate and the quantum mechanical one is too time-consuming and/or complicated. The model especially can be seen as a refinement to B. Grosswendts model (8, 9) in the sense that the size of the ionisation clusters can be broken down into smaller ones (cf. example given in subsection "Clarification regarding partitions" in Section 2).
The next section is about the model. Then, its numerical realisation will be discussed in Section 3. As the model takes a new approach, a critical assessment of the present work is given in Section 4. And finally, a summary and outlook are presented at the end in Section 5.

## 2. Model

Recall that the spatial extension of the wave packet usually is larger than the target volume for low-energy (i. e. $\leq 100$ eV) particles moving in nanometric volumes. So, the classical description, which is based on trajectories, has come to a limit. The appropriate description would have to be within the framework of quantum mechanics. The explicit calculations, however, would be too sophisticated and time-consuming. The model that follows is an attempt to do it better than the classical description and quicker than the quantum mechanical one, although the latter would be best, of course. The model is a statistical ionisation-clustering model. It is intended as an alternative proposal and still offers room for improvement. It is based on the following scenario and assumptions: starting point is a nanometric target volume consisting of liquid water (a DNA polymer could also be used instead of liquid water), which can have any shape. The volume is irradiated with directly ionising (primary) particles which could have low energy in particular. The target and the primary particles (including source) can be located in a vacuum or embedded in a medium (e.g., air). The primary particles can be electrons or positrons. We will focus on electrons below. An illustration of this scenario is shown in Figure 2.

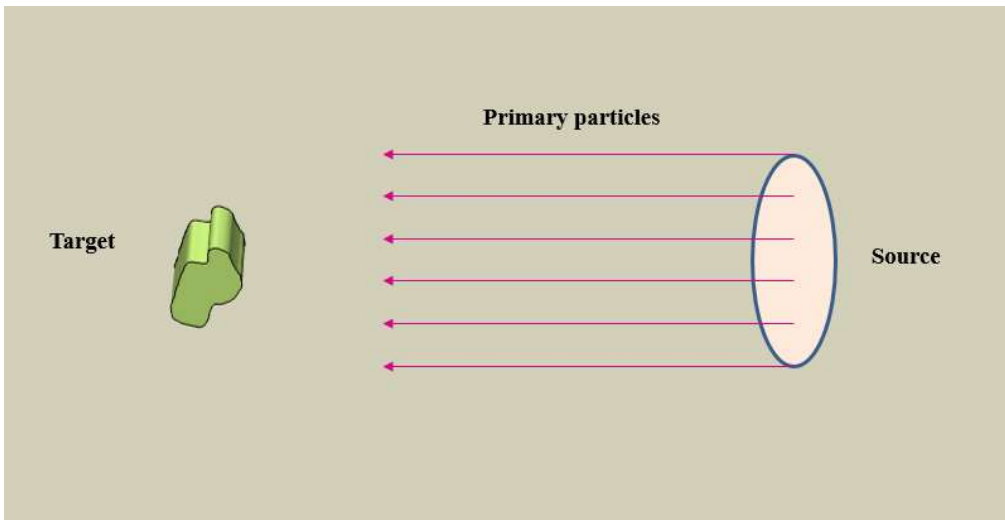

*Figure 2: exemplary illustration of an underlying scenario. The target can be arbitrarily shaped. The primary particles are ionising. Without loss of generality, a plane-parallel, homogeneous source was drawn. Target, primary particles, and source can be embedded in a medium (e.g., air).*

Ionisations can occur at energies of around 10 eV due to the interactions of the primary particles as well as secondary electrons (and higher-generation electrons, respectively) with the material (even at lower energies, i.e. in the range of a few meV, changes in the molecular structure may already occur). In addition, we assume:

- i.) The ionisations and the subsequent 'rearrangement processes' (for example, electron hopping, electron attachment, and recombination (30)) in the target volume lead to an order-disorder phase transition, whereby the statistics dominate (due to the large number of degrees of freedom being involved) the dynamics (which become very complex at the phase transition; correlation length diverges). This means that the vertices for ionisation and excitation generally do not coincide with the final ion positions (or positions of DNA breaks).
- ii.) The totality of the sizes of the individual clusters of ionisations present after these actions (cf. point i.) can be described by a stochastic distribution based on the free energy.
- iii.) The statistical ensemble of imaginary uniform ionisation-cluster systems is a canonical one. Note: A microcanonical description does not seem appropriate since the energy is not conserved. A grand-canonical description does not seem appropriate either, because the total number of ionisations within the target does not alter.

We mostly make use of the following nomenclature:

$n_t$: total number of ionisations; it is in some sense the nanometric analogue of $D_{abs}$ (its mean value correlates to $D_{abs}$)

q(I): Number of ways to write or decompose an integer "I" as a sum. For example, if I = 4 then q(4) = 5 since 4 = 4 + 0 [*1<sup></sup>st option*] = 3 + 1 [*2nd option*] = 2 + 2 [*3rd option*] = 2 + 1 + 1 [*4th option*] = 1 + 1 + 1 + 1 [*5th option*]. The integers can be understood as size of an ionisation cluster. Therefore, q(I) also indicates the number of possible partitions into which a given number of ionisations can be decomposed, whereat a partition is a set of ionisation clusters (see below; in particular the section "Clarification regarding partitions") and represents a microstate.

$n_j$: number of j interrelated ionisations. We call j interrelated ionisations an "ionisation cluster of size j". Its probability of occurrence is related to the free energy. The following applies for one partition: $j \in [1, ..., k]$ and $n_k \leq n_t$.

$\overrightarrow{n_t}$: t-dimensional partition vector, $(n_1, n_2, ..., n_t)^t$, describing one of the $q(n_t)$ partitions; in a sense, it is the nanometric analogue to the lineal energy. Please note: a partition vector is not a "real" vector in terms of Euclidean geometry.

$\{\overrightarrow{n_t}\}$: set of all possible $q(n_t)$ t-dimensional partition vectors

M: ionisation-cluster multiplicity of a partition, i. e. number of ionisation clusters of a partition

$(n_t, M)$: set of partitions which refer to $n_t$ ionisations and M ionisation clusters

$q(n_t|M)$: number of partitions of $n_t$ with ionisation-cluster multiplicity M, i. e. cardinality of $(n_t, M)$
Note: $q(n_t|M)$ is not equal to the conventional partition sum.

L: likelihood of $(n_t,M)$

V: nanometric volume

T: temperature

event: consists of a primary particle or higher-generation electron (which is able to ionise) moving through V

F: free energy

$\varepsilon_0$: vacuum permittivity

$k_B$: Boltzmann constant

Clarification regarding partitions:
If there are $n_t = 5$ ionisations, e. g., then there are q(5) = 7 partition vectors $\overrightarrow{n_{5,i}}$ (i = 1, ..., 7).
This means, $\{\overrightarrow{n_5}\} = \{ \overrightarrow{n_{5,1}} = (5, 0, 0, 0, 0)^t, \overrightarrow{n_{5,2}} = (3, 1, 0, 0, 0)^t, \overrightarrow{n_{5,3}} = (1, 2, 0, 0, 0)^t, \overrightarrow{n_{5,4}} = (2, 0, 1, 0, 0)^t, \overrightarrow{n_{5,5}} = (0, 1, 1, 0, 0)^t, \overrightarrow{n_{5,6}} = (1, 0, 0, 1, 0)^t, \overrightarrow{n_{5,7}} = (0, 0, 0, 0, 1)^t \}$. The partition vector $\overrightarrow{n_{5,4}}$ refers to two ionisation clusters of size 1 and to one ionisation cluster of size 3. There are q(5|3) = 2 partition vectors with ionisation-cluster multiplicity M = 3 (which belong to (5, 3)): $\overrightarrow{n_{5,3}}$ and $\overrightarrow{n_{5,4}}$.
The above example illustrates that in our model, five ionisations in the target volume can lead not only to one ionisation cluster of size 5 — as in B. Grosswendt (8) — but also to six further ionisation-cluster formations: one ionisation cluster of size 2 and one of size 3 (represented by $\overrightarrow{n_{5,5}}$), one ionisation cluster of size 1 and two of size two (represented by $\overrightarrow{n_{5,3}}$), and so on.

The basic idea of the model is as follows – its implementation is outlined in the next section.
For all events do:

- calculate all possible partitions of the number of ionisation interactions $n_t$,
- calculate the temperature T and the free energy F, and
- choose a set of partitions which refers to $n_t$ ionisations and M ionisation clusters due to "$\exp\{-F((n_t, M), T, V)/T\}$" as well as select a microstate with equal probability then.

## 3. Simulation Methodology

The methodology reads as follows in detail:

**(1) Numerical calculation of $n_t$**
To calculate the total number of ionisations, $n_t$, a scoring volume with the shape of a cylinder is preconditioned (without loss of generality). The material inside the cylinder is DNA or DNA-equivalent matter (which, for example, may be liquid water for the time being) as mentioned above. The numerical calculation is done using a classical particle transport code (like Geant4-DNA (13, 14, 24 – 26), MDM (15), or PARTRAC (16); an overview of common classical particle transport codes can be found in (17)). We presume here that $n_t$ will not change much for a quantum multiple-scattering calculation since ionising interactions are inelastic ones, i. e. we suppose a circumstantial validity of the trajectory method (28).

**(2) Computation of $\{\overrightarrow{n_t}\}$ and selection of $\overrightarrow{n_t}$**

The number of partitions of $n_t$ can become very large, as its calculation is a so-called NP-complete decision problem: if $n_t = 10$, then q(10) = 42; if $n_t = 100$, then q(100) = 190569292; and if $n_t = 1000$, then q(1000) = 24061467864032622473692149727991; see (18).

If the number of partitions is small, i. e. if, for example, all microstates may be calculated within 2 hours cpu time, then we take into account all partition vectors and proceed as described in subsection (2.1) below. If the number of partitions is not small, we propose to consider "biased" subsets of the multidimensional partition space (19) and then apply the procedure of subsection (2.1). In future, however, it may be possible to store and compute partitions using quantum registers and quantum gates. This would enable simulations based on a large number of partitions to be carried out using a computer whose central processing unit has been expanded to include a quantum processing unit.

In many cases, however, $n_t$ should be less than 100 – and the calculation for this is almost instantaneous.

**(2.1) Calculation of likelihood L**

To clarify the structure of the calculation of the likelihood L, we divide this section into two sub-sections, termed (2.1.1) and (2.1.1.1), where the subsection (2.1.1.1) has to be processed before the subsection (2.1.1). For the calculation of L, we assume that the systems of imaginary uniform ionisation-clusters can be described within the canonical approximation. Using the maximum entropy principle (29), the likelihood of $(n_t, M)$ is then given by

$$L = c * \exp\{-F((n_t, M), T, V)/T\}, \tag{1}$$

where "c" is a normalisation constant into which the absorbed energy (cf. Equation 18) and the temperature "T" are introduced, and "F" is the free energy of the system being in thermal equilibrium. The nanometric target volume is designated "V". Having chosen a set of partitions that refers to $n_t$ ionisations and M ionisation clusters, one microstate, described by $\overrightarrow{n_{t,i}}$ (with i ϵ {1, ..., $q(n_t|M)$}), is selected with equal probability then. (The assumption of equal probability is a simplifying approximation. In principle, one could discard this assumption and calculate the probability for each individual partition (microstate)).

**(2.1.1) Calculation of free energy F**

The free energy "F" of the macroscopic system of ionisation clusters is assumed to be the sum of the "translational" parts "$F_j^{tr,Z}$" and the "inner" parts "$n_j F_j^{in}$",

$$F((n_t, M), T, V) = \sum_{j=1}^{n_t} (F_j^{tr,Z} + n_j F_j^{in}), \tag{2}$$

with "$n_j$" being the number of ionisation clusters of size "j" (remark: there is no factor $n_j$ missing before "$F_j^{tr,Z}$" , as $n_j$ is already contained in "$F_j^{tr,Z}$"; cf. Equations 4 and 14). Each $n_j$ contains the contributions of the corresponding microstates:

$$n_j = \sum_{k=1}^{q(n_t|M)} n_{j,k}, \tag{3}$$

where "$n_{j,k}$" refers to the number of ionisation clusters of size "j" of microstate "k".

For the explicit calculation of the translational part "$F_j^{tr,Z}$" of the free energy, we assume that the ionisation clusters behave in the same way as particles of a Boltzmann gas (this may be a rough approximation as regards DNA — similar to the approximation in which DNA is approximated with water.), i. e. their spatial distribution is essentially uniform. We assume further that the ionisation clusters are within a cylindrical target volume V. Furthermore, the Coulomb interaction is treated by means of the mean-field approximation (whereat ionisations outside the target are neglected).

Let $\vec{x}$ contain the coordinates of all $n_j$ ionisation clusters of size j. The canonical partition sum then reads,

$$Z(T, n_j, V) = \frac{\lambda_T^{-3n_j}}{n_j!} \int e^{-\beta V_m(\vec{x})} \, d^{3n_j}x \, . \tag{4}$$

The integration is over volume V. The abbreviation $\lambda_T$ indicates the thermal de Broglie wavelength,

$$\lambda_T = \frac{h}{\sqrt{2\pi m k_\beta T}} \ , \tag{5}$$

where "h" denotes Planck's constant and "m" is the mass of the ionisations clusters of size "j". The mass is obtained by multiplying "j" by the respective molecular mass. The factor "$k_B$" is the Boltzmann constant, and the thermodynamic beta reads, $\beta = 1/(k_B T)$. In Equation 4, the kinetic energy term of the Hamiltonian (which implicitly is incorporated in the exponential function) is contained in the leading factor. The potential energy term, which is due to the Coulomb field, is denoted by "$V_m$". All ionisation clusters are exposed to the same field. Hence, we consider $V_m(\overleftarrow{x})$ as the sum of the contributions of the individual clusters, $V_m(\overleftarrow{x}) = V'_m(\overleftarrow{x_1}) + \cdots + V'_m\left(\overleftarrow{x_{n_j}}\right)$, where each contribution $V'_m(\overleftarrow{x_i})$ $(i \in \{1, \cdots, n_j\})$ refers to the Coulomb energy of a cluster of size "j" at position $\overleftarrow{x_i}$ (inside the target volume) in the field of the remaining clusters and reads,

$$V'_m(\overleftarrow{x}_i) = V'_m(x_i) = \frac{e^2 j(n_t - j)}{8\pi\varepsilon_0 R}\left(3 - \frac{x_i^2}{R^2}\right). \tag{6}$$

The vacuum permittivity is abbreviated to "$\varepsilon_0$" and "e" is the elementary charge. In Equation 6, it was also assumed for simplicity that the target volume is a sphere and that the total charge of the remaining clusters is distributed homogeneously within this sphere. The radius R of the sphere is determined so that the sphere has the same volume as the cylindrical target volume (the difference between the cylindrical and the spherical shape is neglected, which seems justified due to the small dimensions). After performing the integration, this term is obtained for the canonical partition sum:

$$Z(T, n_j, V) = \frac{\lambda_T^{-3n_j}}{n_j!} \left(C_1 C_2 e^{-\beta H_1}\right)^{n_j} , \tag{7}$$

with

$$C_1 = 2\,\pi\,R\,k^{-1} \ , \tag{8}$$

$$C_2 = e^{kR^2} - \sqrt{\frac{\pi}{k}} \cdot \frac{1}{2\,R} \cdot erfi(R \cdot \sqrt{k}) \text{ , where} \tag{9}$$

"erfi" denotes the imaginary error function,

$$H_1 = \frac{3\,e^2 j(n_t - j)}{8\pi\varepsilon_0 R} \ , \tag{10}$$

$$k = \beta\,H_2 \ , \tag{11}$$

$$\beta = \frac{1}{k_B T} \quad \text{and} \tag{12}$$

$$H_2 = H_1\,\frac{3}{R^2} \ . \tag{13}$$

Finally, we find for the translational part "$F_j^{tr,Z}$":

$$F_j^{tr,Z} = -\, k_B T \ln Z(T, n_j, V). \tag{14}$$

The term "$F_j^{in}$" of the inner parts is obtained using a generalisation of the well-known nuclear droplet model. For the time being, we consider the Coulomb contribution only for the calculation of $F_j^{in}$. Starting point is the Coulomb energy $E_j^C$ of "j" charges which are homogeneously distributed within a sphere of radius "$\tilde{r}_k$":

$$E_j^C = \frac{1}{4\pi\varepsilon_0} \frac{3}{5} \frac{e^2 j^2}{\tilde{r}_k}. \tag{15}$$

The radius "$\tilde{r}_k$" is obtained from the requirement that the number of charges per sphere volume corresponds to the molecular density "ρ" in the target volume:

$$\frac{j}{\frac{4\pi}{3}\tilde{r}_k^3} = \rho \;\; \Rightarrow \;\; \tilde{r}_k = \sqrt[3]{3\, j/(4\,\pi\,\rho)} \; . \tag{16}$$

Furthermore, we assume $E_j^C$ to be (almost) independent of the temperature. Therefore, there is only one microstate which is associated with this energy. The respective entropy, hence, is zero. So, we can write

$$F_j^{in} = E_j^C. \tag{17}$$

**(2.1.1.1) Calculation of temperature T**

The temperature T is obtained from the absorbed energy "$E_{abs}$", which is described by

$$E_{abs} = \frac{3\, e^2 n_t^2}{20\, \pi \varepsilon_0\, R} + \sum_{j=1}^{n_t} n_j \left[F_j(T,V) - T\left(\frac{\partial F_j(T,V)}{\partial T}\right)_V\right] - E_0. \tag{18}$$

The first term on the right hand side refers to the Coulomb energy of a homogeneously charged sphere of radius R. Then the internal energy follows. The term "$E_0$" denotes the ground-state energy. We approximate $E_0$ by the corresponding expression for an ideal gas,

$$E_0 = \frac{3}{2}\, n_t\, k_B T_{initial} \,, \tag{19}$$

where "$T_{initial}$" denotes the initial temperature. The absorbed energy can be computed by one of the classical particle transport codes mentioned above (see subsection **(1)**); just as $n_t$. The information contained in the free energy of Equation 18 is also contained in the absorbed energy. There is therefore no chicken-and-egg problem with respect to F and T, as one might assume with regard to Equation 18. After some tedious (albeit elementary) calculations, Equation 18 can finally be transformed into a simple term for the temperature:

$$T = \Theta_1 E_{abs} - \Theta_2 \,. \tag{20}$$

Factor $\Theta_1$ is given by

$$\Theta_1 = \left[\frac{3}{2}\, k_B \left(\sum_{j=1}^{n_t} n_j^2 - n_t\right)\right]^{-1}, \tag{21}$$

and term $\Theta_2$ reads:

$$\Theta_2 = \Theta_1 \frac{3\, e^2}{20\, \pi\, \varepsilon_0\, R} \left( n_t^2 + \sum_{j=1}^{n_t} \left(\frac{R}{\tilde{r}_k} + n_j^2\right) j^2 n_j \right) . \tag{22}$$

As an alternative to Equation 18, one could also calculate the temperature using the Saha equation (21) or using the specific heat capacity (22) for the temperature increase.

**(3) Sorting ionisation clusters into a histogram**
This is done as usual.

**(4) Repeating Nos 1 to 3 until the end is reached**
The end is determined by the total number of events.

## 4. Critical Assessment

The model described above is an attempt to overcome the limitations of clustering models, which are due to the situation that the classical trajectory approach is no longer adequate (for light, charged, low-energy primary leptons) on the one hand and a "free" or "to some extent arbitrarily chosen" parameter exists on the other hand (cf. introduction). In this respect, it is intended in particular to offer an alternative to previous practice, in which classical models and software packages, respectively, are applied to the case "low-energy (< 100 eV) primary electrons moving in nanovolumes (in the order of a few nanometers)". Furthermore, in contrast to previous clustering models, the model allows statements to be made about the thermodynamic state quantities entropy and free energy, respectively, and the temperature. In addition, the model could also be tailored relatively easily to a grand-canonical ensemble or a micro-canonical one if this would be more suitable for another scenario. The model therefore offers a relatively large potential of application (in particular, it is well suited for the transition from nanodosimetry to microdosimetry). However, only the basic concept was presented in Section 2. The model still provides ample scope for single improvements, e.g. with regard to the calculation of the free energy and the temperature. Therefore, the model still needs to be evaluated thoroughly and carefully. This is particularly important with regard to the assumption of the order-disorder phase transition – which may well be the most critical assumption of the model. A thorough and careful evaluation will be subject of a forthcoming paper.

## 5. Summary and Outlook

A proposal on the calculation of the ionisation-cluster size distribution was depicted. The proposal was inspired by the maximum entropy principle (29) and the well-known nuclear droplet model (19, 29). In contrast to previous models, our model allows statements about thermodynamic state quantities and it can be applied in particular to the scenario of "low-energy (< 100 eV) primary electrons moving in nanovolumes (in the order of a few nanometers)" – a scenario for which trajectory models should not be used. However, much work remains to be carried out for a proper validation! This applies in particular to the calculation of the free energy and the temperature as well as to the assumed order-disorder phase transition. The labour-intensive work that still needs to be done will be subject of a forthcoming paper. (A comparison with algorithms for DNA clustering into radiation-induced foci, cf. (23), also seems worthwhile.) The focus of this work was on presenting the model.

## Acknowledgements

The author is deeply grateful to Professor C. R. Arumainayagam for information on the accuracy of electron guns. The author also gratefully acknowledges the support by the NUSAFE programme.

## References

1. Rossi HH and Zaider M. Microdosimetry and Its Applications. Berlin: Springer (1996).

2. Cordoni F, Missiaggia M, Attili A, Welford SM, Scifoni E, and La Tessa C. Generalized stochastic microdosimetric model: The main formulation. Phys. Rev. E (2021) 103, 012412.
3. Friedland W, Jacob P, Paretzke HG, and Stork T. Monte Carlo Simulation of the Production of Short DNA Fragments by Low-Linear Energy Transfer Radiation Using Higher-Order DNA Models. Radiation Research (1998) 150, p. 170 – 182.
4. Besserer J and Schneider U. A track-event theory of cell survival. Z. Med. Phys. (2015) 25, 168.
5. Besserer J and Schneider U. Track-event theory of cell survival with second-order repair. Radiat. Environ. Biophys. (2015) 54, 167.
6. Schneider U, Vasi F, Schmidli K, and Besserer J. A model of radiation action based on nanodosimetry and the application to ultra-soft X-rays. Radiat. Environ. Biophys. (2020) 59, 439.
7. Ngcezu SA and Rabus H. Investigation into the foundations of the track-event theory of cell survival and the radiation action model based on nanodosimetry. Radiat. Environ. Biophys. (2021).
8. Grosswendt B. NANODOSIMETRY, FROM RADIATION PHYSICS TO RADIATION BIOLOGY. Radiation Protection Dosimetry (2005) Vol. 115, p. 1 – 9.
9. Grosswendt B. RECENT ADVANCES OF NANODOSIMETRY. Radiation Protection Dosimetry (2004) Vol. 110, p. 789 – 799.
The model goes back to: De Nardo L, Colautti P, Conte V, Baek WY, Grosswendt B, and Tornielli G. Ionization-cluster distributions of α-particles in nanometric volumes of propane: measurement and calculation. Radiat. Environ. Biophys. (2002) 41, 235–256.
10. Villegas F, Bäckström G, Tilly N, and Ahnesjö A. Energy deposition clustering as a functional radiation quality descriptor for modeling relative biological effectiveness. Med. Phys. (2016) 43 (12).
11. Garty, G., Schulte, R., Shchemelinin, S., Leloup, C., Assaf, G., Breskin, A., Chechik, R., Bashkirov, V., Milligan, J., and Grosswendt, B., *A nanodosimetric model of radiation-induced clustered DNA damage yields*, Med. Phys. (2010) 55, p. 761–781.
12. Liljequist, D., A study of errors in trajectory simulation with relevance for 0.2– 50 eV electrons in liquid water, Radiation Physics and Chemistry (2008) 77, p. 835 – 853.
13. Incerti, S., Baldacchino, G., Bernal, M., Capra, R., Champion, C., Francis, Z., Guatelli, S., Guèye, P., Mantero, A., Mascialino, B., et al., *The Geant4-DNA project*, Int. J. Model. Simul. Sci. Comput. (2010) 1, p. 157 – 178.
14. Bernal MA, Bordage MC, Brown JMC, Davídková M, Delage E, El Bitar Z, Enger SA, Francis Z, Guatelli S, Ivanchenko VN, et al. Track structure modeling in liquid water: A review of the Geant4-DNA very low energy extension of the Geant4 Monte Carlo simulation toolkit. Phys. Med. (2015) 31, p. 861 – 874.
15. Poignant F, Ipatov A, Chakchir O, Lartaud P-J, Testa É, Gervais B, and Beuve M. Theoretical derivation and benchmarking of cross sections for low-energy electron transport in gold. Eur. Phys. J. Plus (2020) 135: 358.
16. Paretzke HG, Turner JE, Hamm RN, Wright HA, Ritchie RH. Calculated yields and fluctuations for electron degradation in liquid water and water vapor, J. Chem. Phys. (1986) 84, p. 3182 – 3188.
17. Li WB, Belchior A, Beuve M, Chen YZ, Di Maria S, Friedland W, Gervais B, Heide B, Hocine N, Ipatov A, Klapproth AP, Li CY, Li JL, Multhoff G, Poignant F, Qiu R, Rabus H, Rudek B, Schuemann J, Stangl S, Testa E, Villagrasa C, Xie WZ, and Zhang YB. Intercomparison of Dose Enhancement Ratio and Secondary Electron Spectra for Gold Nanoparticles Irradiated by X-rays Calculated Using Multiple Monte Carlo Simulation Codes. Physica Medica (2020) 69, p. 147-163. And corrigendum:

Li WB, Beuve M, Di Maria S, Friedland W, Heide B, Klapproth AP, Li CY, Poignant F, Rabus H, Rudek B, Schuemann J, and Villagrasa C. Corrigendum to "Intercomparison of Dose Enhancement Ratio and Secondary Electron Spectra for Gold Nanoparticles Irradiated by X-rays Calculated Using Multiple Monte Carlo Simulation Codes" [Phys. Med. 69 (2020), p. 147-163]. Physica Medica (2020) 80, p. 383–388.

18. Bruinier JH and Ono K. Algebraic formulas for the coefficients of half-integral weight harmonic weak Maass forms. arXiv:1104.1182 [math.NT], https://doi.org/10.48550/arXiv.1104.1182 [Accessed March 21, 2024].
19. Bondorf JP, Botvina AS, Iljinov AS, Mishustin IN, and Sneppen K. STATISTICAL MULTIFRAGMENTATION OF NUCLEI. Physics Reports (1995) 257 p. 133 – 221.
20. Bondorf JP, Donangelo R, Mishustin IN, Pethick CJ, Schulz H, and Sneppen K. STATISTICAL MULTIFRAGMENTATION OF NUCLEI, (I). Formulation of the model. Nuclear Physics (1985) A443, p. 321 – 347.
21. Chen FF. Introduction to Plasma Physics and Controlled Fusion. Springer International Publishing Switzerland (2018), ISBN 978-3-319-22308-7.
22. https://phys.libretexts.org/Bookshelves/University_Physics/Physics_(Boundless)/13%3A_Heat_and_Heat_Transfer [Accessed September 25, 2024].
23. Poignant F, Pariset E, Plante I, Ponomarev AL, Evain T, Viger L, Slaba TC, Blattnig SR, and Costes SV. DNA break clustering as a predictor of cell death across various radiation qualities: influence of cell size, cell asymmetry, and beam orientation. Integrative Biology (2024), 16, zyae015, https://doi.org/10.1093/intbio/zyae015 [Accessed September 30, 2024].
24. Tran HN, Archer J, Baldacchino G, Brown JMC, Chappuis F, Cirrone GAP, Desorgher L, Dominguez N, Fattori S, Guatelli S, Ivantchenko V, Méndez J-R, Nieminen P, Perrot Y, Sakata D, Santin G, Shin W-G, Villagrasa C, Zein S, Incerti S. Review of chemical models and applications in Geant4-DNA: Report from the ESA BioRad III Project. Med. Phys. (2024) 51, 5873–5889.
25. Incerti S, Kyriakou I, Bernal MA, Bordage MC, Francis Z, Guatelli S, Ivanchenko V, Karamitros M, Lampe N, Lee SB, Meylan S, Min CH, Shin WG, Nieminen P, Sakata D, Tang N, Villagrasa C, Tran H, Brown JMC. Geant4-DNA example applications for track structure simulations in liquid water: a report from the Geant4-DNA Project. Med. Phys. (2018) 45, e722–e739.
26. Incerti S, Ivanchenko A, Karamitros M, Mantero A, Moretto P, Tran HN, Mascialino B, Champion C, Ivanchenko VN, Bernal MA, Francis Z, Villagrasa C, Baldacchino G, Guèye P, Capra R, Nieminen P, Zacharatou C. Comparison of Geant4 very low energy cross section models with experimental data in water. Med. Phys. (2010) 37, 4692–4708.
27. Arumainayagam, C. R., private communication. See also ELG-2A-6214/EGPS-1022D-1129 User Manual (2011), Kimball Physics, Inc., Wilton, New Hampshire 03086, USA.
28. Liljequist D and Nikjoo H. On the validity of trajectory methods for calculating the transport of very low energy (< 1 keV) electrons in liquids and amorphous media. Radiation Physics and Chemistry (2014) 99, p. 45 – 52.
29. Jaynes ET. Information Theory and Statistical Mechanics. Phys. Rev. (1957) 106, p. 620 – 630.
30. Gao Y, Zheng Y, and Sanche L. Low-Energy Electron Damage to Condensed-Phase DNA and Its Constituents. Int. J. Mol. Sci. (2021), 22, 7879.